\documentclass[a4paper,journal]{IEEEtran}
\usepackage{amsmath,amsfonts}
\usepackage[linesnumbered,ruled,vlined]{algorithm2e}  
\usepackage{algorithmicx}
\usepackage{algcompatible}
\usepackage{array}
\usepackage{textcomp}
\usepackage{stfloats}
\usepackage{url}
\usepackage{verbatim}
\usepackage{graphicx}
\usepackage{cite}
\usepackage{makecell, booktabs, multirow}
\usepackage{enumitem}
\usepackage{lipsum}
\usepackage{xcolor}
\usepackage{caption}
\usepackage{subcaption}
\newcommand{\ket}[1]{\vert #1 \rangle}

\begin{document}
\bstctlcite{IEEEexample:BSTcontrol}

\title{\huge Scalable Quantum Message Passing Graph Neural Networks for Next-Generation Wireless Communications: Architectures, Use Cases, and Future Directions}

\author{Le Tung~Giang, Nguyen Xuan~Tung, Trinh Van~Chien,~\IEEEmembership{Member,~IEEE} \\ Lajos~Hanzo, ~\IEEEmembership{Life Fellow,~IEEE}, and Won-Joo~Hwang,~\IEEEmembership{Senior Member,~IEEE}
\thanks{Le Tung~Giang is with the Department of Information Convergence Engineering, Pusan National University, Busan 46241, Republic of Korea (e-mail: giang.lt2399144@pusan.ac.kr).}
\thanks{Nguyen Xuan Tung is with the Faculty of Interdisciplinary Digital Technology, PHENIKAA University, Yen Nghia, Ha Dong, Hanoi 12116, Viet Nam (e-mail: tung.nguyenxuan@phenikaa-uni.edu.vn). }
\thanks{Trinh Van~Chien is with the School of Information and Communications Technology, Hanoi University of Science and Technology, Hanoi 100000, Vietnam (e-mail: chientv@soict.hust.edu.vn).}
\thanks{Lajos Hanzo is with the Department of Electronics and Computer Science, University of Southampton, Southampton SO17 1BJ, U.K. (e-mail: lh@ecs.soton.ac.uk).}
\thanks{Won-Joo Hwang is with the School of Computer Science and Engineering, Center for Artificial Intelligence Research, Pusan National University, Busan 46241, South Korea (e-mail: wjhwang@pusan.ac.kr).}
\thanks{\textit{(Corresponding author: Won-Joo Hwang)}}
}



\maketitle

\begin{abstract}
Graph Neural Networks (GNNs) are eminently suitable for wireless resource management, thanks to their scalability, but they still face computational challenges in large-scale, dense networks in classical computers.
The integration of quantum computing with GNNs offers promising pathway for enhancing computational efficiency, because they reduce the model complexity. This is achieved by leveraging the quantum advantages of parameterized quantum circuits (PQCs), while retaining the expressive power of GNNs.
However, existing pure quantum message passing models remain constrained by the limited number of qubits, hence limiting the scalability of their application to the wireless systems.  
As a remedy, we conceive a Scalable Quantum Message Passing Graph Neural Network (SQM-GNN) relying on a quantum message passing architecture. To address the aforementioned scalability issue, we decompose the graph into subgraphs and apply a shared PQC to each local subgraph. 
Importantly, the model incorporates both node and edge features, facilitating the full representation of the underlying wireless graph structure.
We demonstrate the efficiency of SQM-GNN on a device-to-device (D2D) power control task, where it outperforms both classical GNNs and heuristic baselines. These results highlight SQM-GNN as a promising direction for future wireless network optimization.

\end{abstract}


\section{Introduction}

\begin{figure*}[t]
    \centering
    \includegraphics[width=\linewidth]{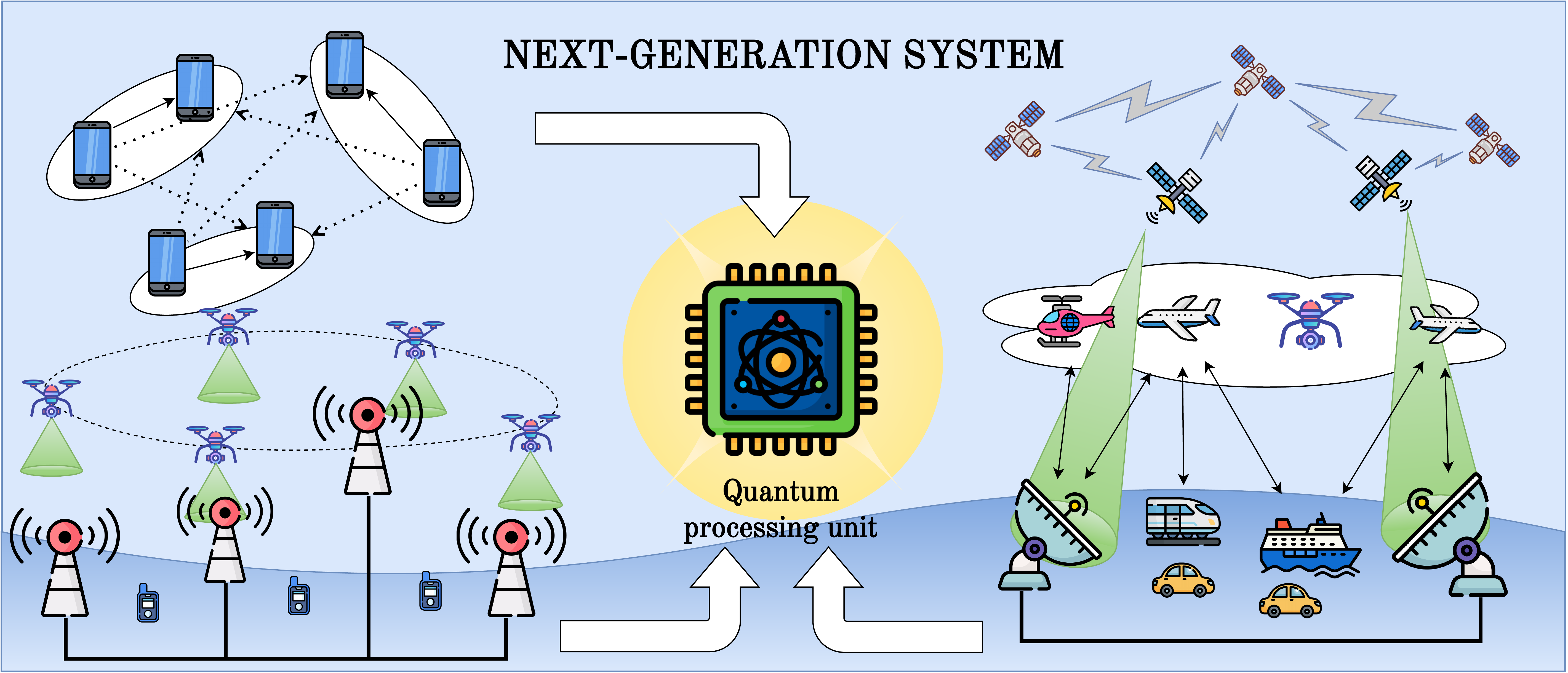}
    \caption{An overview of QPU-assisted NG communication architecture, where the QPU serves as a central resource management entity, connecting and coordinating multiple platforms.}
    \label{fig:6G_applications}
    \vspace{-0.75cm}
\end{figure*}

\IEEEPARstart{N}{ext} generation (NG) wireless networks are expected to support dense deployments, decentralized architectures, and highly dynamic environments. NG infrastructures will span diverse spatial domains across air, ground, and space. This includes cell-free massive MIMO (cf-mMIMO), UAV-assisted relays, and satellite-based systems,  leading to large-scale, heterogeneous, and graph-structured communication topologies, as illustrated in Fig.~\ref{fig:6G_applications}. Traditional optimization methods, while effective in fixed scenarios, often falter when it comes to handling the dynamic behavior and flexibility required by modern systems. 
To support real-time operation, deep learning (DL) techniques have attracted research attention for their ability to model nonlinear system dynamics and deliver prompt inference. Among them, graph neural networks (GNNs) have shown strong potential in wireless communications due to their capability to capture spatial and topological dependencies~\cite{2024_Tung_GNNSurvey}. GNNs are naturally suited to network systems, offering permutation equivariance, strong generalization across diverse network sizes, and scalability in both centralized and distributed architectures~\cite{2024_SensJ_Giang_HGNN, 2024_TVT_Tung_GNN_cfmMIMO}. 
However, as networks grow in scale, density and connectivity, GNNs encounter challenges such as over-smoothing in deep architectures, and classical computing bottlenecks, limiting the scalability to moderate-dimensional graphs~\cite{2017_NIPS_GraphSAGE}.

\textcolor{black}{
To overcome these limitations, quantum machine learning (QML) has recently emerged as a promising technique of enhancing classical learning models. 
By leveraging the unique features of quantum computing such as entanglement, superposition, and quantum parallelism, QML that can accelerate convergence, reduce model complexity, and offer a more expressive representation space. Recent works have already demonstrated the potential of QML for wireless communication problems, achieving performance beyond what is attainable with purely classical approaches~\cite{2024_TWC_QML}. 
}
At the core of most QML models are parameterized quantum circuits (PQCs), which are trainable quantum-domain counterparts of learning modules in classical DL models. PQCs provide a structured way to representing and manipulating information in quantum states, offering advantages in both efficiency and representation. 
Several studies have shown that QML models may outperform their classical counterparts in high-dimensional settings.
This is because, owing to the superposition principle, quantum gates provide their responses to all legitimate inputs at once, while classical gates respond to one at a time.
These strengths suggest that combining QML with GNNs leads to a dedicated Quantum Processing Unit (QPU), as illustrated in Fig.~\ref{fig:6G_applications}. The QPU is expected to coordinate diverse network functions and support dynamic adaptation across heterogeneous environments. It plays a central role in enabling fast, scalable, and topology-aware learning and decision-making in NG scenarios.

Recently, several studies have explored integrating QML into GNNs for enhancing computational capability. 
However, most existing studies adopt a simple hybrid design, where the classical multi-layer perceptrons (MLPs) of GNNs are naively replaced by quantum neural networks (QNNs), leaving the message passing architecture unchanged.
This is the core mechanism in GNNs, where each node aggregates information from its neighbours for updating its representation. Replacing MLPs by QNNs only changes how messages are transformed, but not how they are exchanged or structurally integrated. 
To circumvent this limitation,  the quantum spatial graph convolutional neural network (QSGCN)~\cite{2025_QSGCN_TNNLS} concept directly embeds the message passing mechanism into PQCs, providing quantum-native graph learning.
\textcolor{black}{
However, QSGCN faces significant scalability challenges in the noisy intermediate-scale quantum (NISQ) era. 
Because its qubit and gate requirements grow quadratically with the graph size, QSGCN does not inherit the scalability of classical GNNs. For instance, with around $70$ available qubits, it can only support fully connected graphs of about $12$ nodes, which is far from sufficient for NG wireless scenarios. 
Moreover, although edge-level knowledge of quantities such as channel gains and interference levels is essential for modelling spatial correlations in wireless graphs, QSGCN does not incorporate edge attributes, which further limits its practicality in large-scale wireless applications.
}

We circumvent these limitations by a new architecture intrinsically amalgamating QML with GNN for NG systems, namely by Scalable Quantum Message Passing GNN (SQM-GNN). To the best of our knowledge, this is the first scalable message passing mechanism implemented directly within PQCs for NG wireless communication.
To overcome the NISQ hardware's limitations, we conceive a subgraph decomposition scheme where each local graph is processed by a shared-parameter PQC. 
\textcolor{black}{
By performing graph-structural learning directly within PQCs and operating in a high-dimensional quantum feature space, the proposed design captures richer and inherently more expressive structural dependencies in the graph, while retaining the scalability and permutation equivariance of classical message-passing GNNs.
}
Importantly, our architecture beneficially represents both node features and edge attributes of NG wireless systems.
As a practical validation, we apply the proposed SQM-GNN to a device-to-device (D2D) power control task to demonstrate its clear advantages over classical baselines. These results highlight the potential of SQM-GNNs as a promising direction for intelligent NG resource management. More broadly, this work represents a first step toward quantum-based message passing in communication networks and opens up a new practical path toward inseparably integrating QML and GNNs.

\section{Preliminaries}
In this section, we first lay the foundations of GNNs in wireless communication scenarios, highlighting their current challenges. We then introduce the concepts of quantum computing and QML as promising enhancements to address these limitations and support efficient scalable learning frameworks.

\subsection{Graph Neural Networks in Wireless Communication}

Graphs offer a compact yet expressive representation of communication networks, where nodes correspond to devices and edges capture interactions such as interference or connectivity. GNNs naturally operate based on this structure and are well-suited for learning over variable-sized, topology-aware data in dynamic wireless environments.
Central to their operation is the message passing mechanism, typically structured in three stages: $(i)$ each node shares its current state with neighbours via individual messages, $(ii)$ messages are aggregated, and $(iii)$ the node updates its state based on the aggregated messages. Aggregation and update steps often employ MLPs. After carefully constructed rounds of message passing, the resultant node embeddings are used for downstream-tasks such as classification, regression, or control.

\subsubsection{Wireless Graph Representation}
While node features describe local properties like channel gain or power, edge attributes are essential for modeling pairwise interactions, which fundamentally impact network performance. In wireless GNN applications, incorporating edge features is necessary for accurate and scalable learning.

\subsubsection{Advantages for Wireless Communications}
GNNs have demonstrated excellent performance in wireless resource management tasks such as power control, user association, and beamforming \cite{2024_Tung_GNNSurvey}. Once trained, GNNs provide prompt inference, outperforming classical heuristics while generalizing across diverse network topologies. In contrast to conventional MLPs or CNNs, GNNs exhibit permutation invariance, making them robust to node indexing and scalable to large-scale device population, and antenna arrays. These properties are essential for future ultra-dense, heterogeneous, and dynamic NG networks.

\subsubsection{Challenges for Wireless Communications}
Despite their appealing scalability, GNNs face practical limitations. In dense large-scale graphs, such as those modeling ultra-massive MIMO or cell-free systems, message passing may become computationally expensive and memory-intensive. 
\textcolor{black}{
In principle, a sufficiently expressive GNN can learn the common theoretical structure  shared by many dynamic wireless scenarios. However, achieving this level of expressivity typically requires complex architectures and costly optimization in the classical domain, which becomes a burden in large-scale deployments.
Another key issue in classical GNNs is over-smoothing, where deeper message passing architectures tend to drive node embeddings toward similar values, diminishing the ability to distinguish nodes in large graphs, hence often enforcing the use of overly shallow models.
These challenges motivate sophisticated techniques that amalgamate GNNs with quantum computing, where enhanced message passing can provide more expressive, high-dimensional feature transformation at each layer. This integration allows capturing the underlying physical structure of the wireless network while maintaining shallow depth, thereby mitigating over-smoothing and improving scalability.
}
\vspace{-0.25cm}
\subsection{Quantum Computing Foundations}
Quantum computing offers a unique model of information processing, grounded in principles of quantum mechanics such as superposition, entanglement, and probabilistic measurement. Its core unit, the \textit{qubit}, represents classical bits by mapping them to quantum states that can exist in a superposition of $\vert 0\rangle$ and $\vert 1\rangle$.
These states are manipulated by quantum gates decribed by reversible unitary transformations that evolve the system's state. Single-qubit gates adjust superposition amplitudes, while multi-qubit gates such as controlled NOT (CNOT) induce entanglement, imposing non-classical correlations on qubits. Quantum computation concludes by `observing' or `measuring' its result within a certain measurement basis, which collapses the quantum state into a classical outcome. Although the measurement is probabilistic, carefully designed circuits and observables allow reliable result-extraction of information through repeated trials.

\subsubsection{Quantum Machine Learning}
QNNs combine the representational richness of quantum states with the trainability of PQCs for modeling complex patterns. A typical QNN involves three stages: quantum embedding of classical input, transformation via trainable unitary gates, and measurement of observables to produce classical outputs. 
Given $x \in \mathbb{R}^d$, a quantum feature map $E(x)$ encodes the data into a quantum state, which is evolved under a unitary PQC $U(\boldsymbol{\theta})$. The expectation of a Hermitian observable $H$ over this state yields the model prediction. The model parameters $\boldsymbol{\theta}$ are optimized using variational methods such as the so-called parameter-shift rule, enabling end-to-end training in a hybrid quantum-classical framework.

\subsubsection{Advantages of QML}
QML offers potential advantages for wireless systems, particularly in the context of large-scale optimization. As network density grows, classical resource management faces computational bottlenecks. 
As a remedy, by exploiting superposition and entanglement, QML enables massively parallel information processing across entangled qubits \cite{II_B_2021_TPDS_Singh_A_QComp}. This allows exploration of larger solution spaces and may yield substantial speedups over classical methods. Additionally, recent studies have shown that in high-dimensional regimes, QML often converges faster than its classical counterparts \cite{2025_Hieu_QCNN_cfmMIMO}.

\subsubsection{Challenges of QML}
Despite its promise, the employment of QML faces significant obstacles in wireless networks:
\begin{itemize}[leftmargin=0.5em, itemsep=0.1em, topsep=0.1em]
    \item \textbf{Quantum-Classical Integration:} Most QML models rely on classical optimizers to train their quantum circuits, limiting potential speedups. Moreover, mapping high-dimensional wireless data to quantum states imposes overhead and potential information loss, degrading hybrid model performance.
    \item \textbf{Quantum Hardware:} Current NISQ devices remain limited by short coherence times, restricted qubit counts, and high gate error rates \cite{2018_NISQ_John}. These hardware constraints erode model reliability and stability, underscoring the need for robust quantum error mitigation strategies for practical deployment.
    \item \textbf{Scalability and Trainability:} Deep QNNs often suffer from the \textit{barren plateau} phenomenon, where gradients vanish as the number of qubits increases \cite{II_B_Plateau_Lucas_F}. This severely hampers trainability and generalization in large systems. Addressing this challenge requires designing scalable PQC architectures, decoupling wireless system size from quantum circuit depth.
\end{itemize}

\section{Quantum Message Passing in GNNs}
\label{sec:Method}
\begin{figure*}
    \centering
    \includegraphics[width=\linewidth]{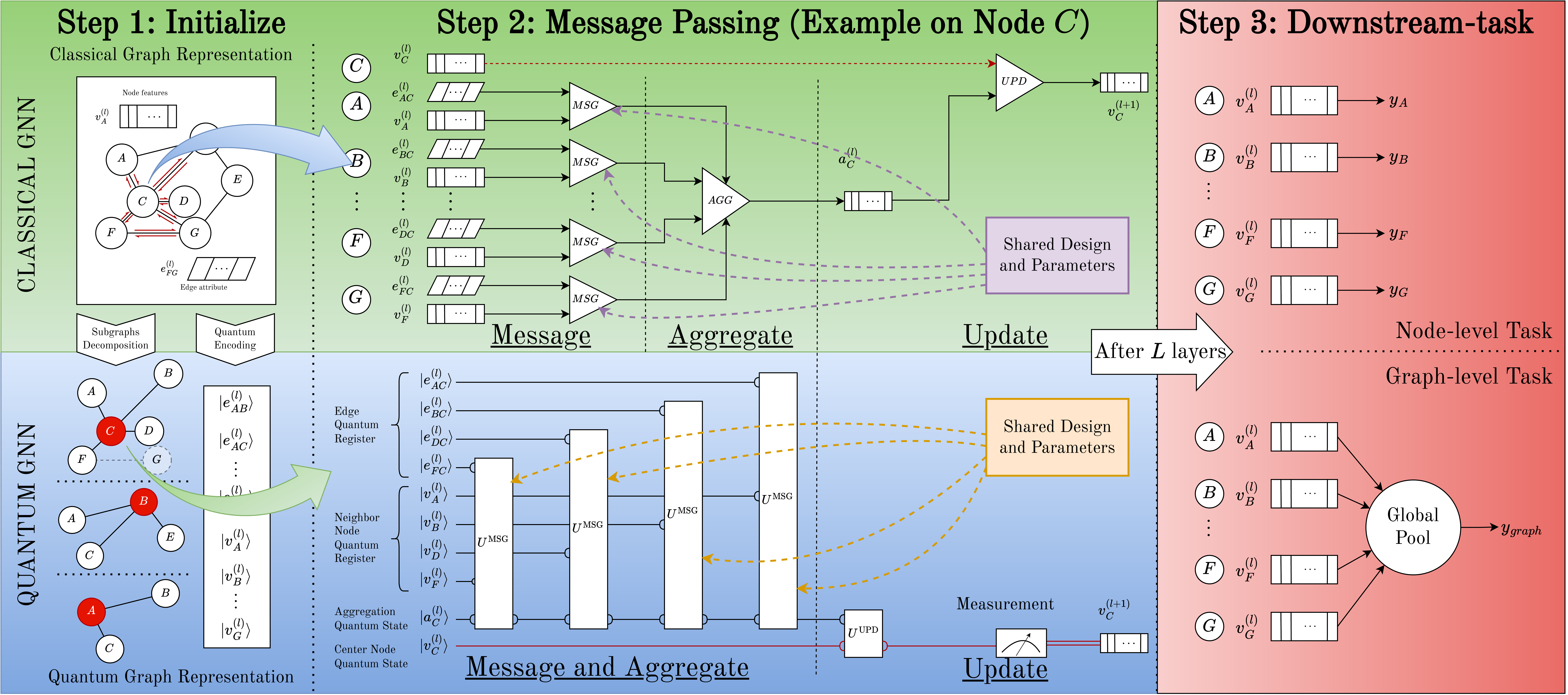}
    \caption{Comparison between the classical GNN architecture (top) and the SQM-GNN architecture (bottom).}
    \label{fig:SQM-GNN_Architecture}
    \vspace{-0.75cm}
\end{figure*}

In this section, we conceive a SQM-GNN architecture that implements the message passing mechanism directly on PQC. Inspired by the classical GNNs, where node features are propagated and aggregated through neighboring connections, the PQC is designed to perform a similar role in the quantum counterpart. The conceptual similarity between classical and quantum message passing is illustrated in Fig.~\ref{fig:SQM-GNN_Architecture}.  

\subsection{SQM-GNN Architecture}
As shown in the top half of Fig.~\ref{fig:SQM-GNN_Architecture}, the classical GNN pipeline proceeds in three stages. First, Step $1$ begins with the initialization of graph data derived from the wireless system. In Step $2$, each node transmits its feature vector with the corresponding edge attribute to neighboring nodes via a shared message-generating function, facilitating the permutation equivariance of GNNs. At the center node $C$, all incoming messages are then aggregated using operators such as mean pooling, attention mechanisms, or neural networks. Finally, in Step $3$, these aggregated messages are then used for updating the central node's embedding.

\textcolor{black}{
As shown in the bottom half of Fig.~\ref{fig:SQM-GNN_Architecture}, SQM-GNN preserves the high-level message passing paradigm of classical GNNs, but implements it via quantum-domain operations. 
By exploiting quantum superposition and entanglement, the QGCL captures richer node and edge interactions at each message passing step, yielding a more expressive structural representation than its classical counterpart.
}
First, in Step $1$, each node feature is encoded into a dedicated qubit in the node quantum registers.
In contrast to prior quantum GNNs, where the edge register merely represents the adjacency, SQM-GNN maps each edge attribute into its own qubit, facilitating a fuller representation of graph structure.
To accommodate the qubit limitations of current NISQ devices, the original graph is decomposed into feasible star‑shaped subgraphs.

In Step $2$, given the list of all sampled subgraphs, each quantum graph convolution layer (QGCL) implements a parameter‑shared message‑passing PQC, consisting of  two stages. 
In particular, first, a unitary $U^{\mathrm{MSG}}$ is applied to each neighbor-edge pair for generating quantum entanglement-based message, represented on the corresponding neighbor-node qubit.
Since the same message unitary is applied uniformly to all neighbor-edge pairs in every subgraph, the message is independent on the node order. Hence, this message generation preserves the permutation equivariant in SQM-GNN.
Then, the message-represented qubits are sequentially amalgamated into center-node qubit via the update unitary $U^{\mathrm{UPD}}$. This imitates the aggregation process in classical GNNs.
Finally, the resulting center‑node state is measured to obtain the refined node embedding, which is forwarded into the next layer.
After several QGCLs, the final node embeddings are forwarded to a classical readout for downstream-tasks, directly analogous to Step $3$ in the classical counterpart. 
The detailed components of this architecture, including quantum encoding, message-passing PQC design, and output handling, are described in the following subsections.

\subsection{Subgraphs Decomposition}
\textcolor{black}{
If we represent each node and edge as separate qubits, the total hardware requirement for realistic wireless graphs would exceed the capabilities of NISQ-era devices.
To accommodate these constraints, each PQC-based QGCL operates on a limited $k$-sampled star-shaped subgraph. This effectively bounds the number of qubits required, while preserving the scalability of the overall SQM-GNN architecture.
}
This subgraph-based approach makes quantum-domain execution feasible on near-term hardware, and also aligns with sampling strategies used in classical GNNs. Empirical studies, such as GraphSAGE~\cite{2017_NIPS_GraphSAGE}, have demonstrated that operating on sampled neighbourhoods can achieve competitive performances while significantly improving runtime and memory efficiency.

This sampling process is illustrated in the lower half of Step~1 in Fig.~\ref{fig:SQM-GNN_Architecture}. For instance, the node $C$ has five neighbours: $\{A,B,D,F,G \}$. In the classical GNN, messages from all neighbours are included during the aggregation and update phases. By contrast, the quantum GNN selectively samples a subset of neighbours to construct the input to the message and aggregation phase. Here, only four out of the five neighbours are selected, the node $G$ has been randomly discarded. Note that, in the next message passing layer, the node $G$ may be randomly selected in the subgraph of node $C$. This ensures that the quantum circuit remains feasible to execute, while still capturing sufficient neighbourhood information.

\subsection{Quantum Graph Convolutional Layer}

\textcolor{black}{
To exploit quantum superposition, entanglement, and parallelism, SQM-GNN performs message passing on each sampled subgraph via PQC-based QGCLs. 
This facilitates a more expressive and robust feature transformation than a pure classical layer, even on the same subgraph structure.
In the following, we first describe how node and edge features are encoded into quantum states, and then explain how message passing is implemented via PQCs.
}

\subsubsection{Quantum Encoding}
In this work, we adopt rotation-based angle encoding as the quantum feature map for embedding classical node and edge features into quantum states. Specifically, each scalar feature is encoded as the rotation angle of a single-qubit gate $R \in \{RX, RY, RZ \}$, resulting in a product state $\ket{\psi} =\otimes^{n}_{i}R(x_i) \ket{0^n}$. This encoding scheme offers multiple advantages over the commonly considered amplitude encoding, which we explicitly avoid for both theoretical and practical reasons. 
Amplitude encoding, while compact in theory, it is primarily simulation-based and impractical on near-term hardware due to its exponentially deep circuits and lack of efficient gate decomposition \cite{2021_PhysRev_Schuld_Encoding}. More critically, amplitude encoding is non-differentiable with respect to the input, making it incompatible with our multi-layer QGCL design, where inputs are re-encoded at each layer. These properties make rotation encoding, rather than amplitude encoding, a practical and principled choice for variational quantum learning in our graph-based framework.

\subsubsection{Message Passing Implementing PQC}
Each QGCL utilizes a shared PQC to implement the message passing mechanism. 
As shown in the lower half of Step $1$ in Fig.~\ref{fig:SQM-GNN_Architecture}, the process begins by encoding the node and edge features into quantum states, forming the Node Quantum Register and Edge Quantum Register, respectively. 
The message passing PQC is implemented using two key unitary operations, $U^{\textrm{MSG}}$ and $U^{\textrm{UPD}}$, which are sequentially executed on the quantum registers, encoding both node features and edge attributes.

For each neighbor-edge pair, the unitary operation $U^{\textrm{MSG}}$ is applied jointly to the neighbor’s node state and the corresponding edge attribute register. This generates a quantum message for each neighbor, embedding both node identity and interaction strength. To prevent destructive interference and preserve the center node’s quantum state, the center node register remains untouched during this phase. 
Once all messages are computed, a second unitary operation $U^{\textrm{UPD}}$ is applied between each quantum state in neighbor node register and the center node quantum state. This operation aggregates quantum messages into the center node through entanglement, effectively combining information from the neighborhood, leveraging quantum computing instead of classical aggregation functions.
Finally, a measurement is then performed on the updated center node quantum state to yield a classical embedding, which is used for updating the center node representation. This quantum message passing process is repeated for each node across multiple layers, allowing the model to capture complex quantum-aware dependencies in the graph structure.
This message passing pipeline is repeated for progressively enriching the node representation learned. Shared PQC structures are reused across nodes, preserving model generalization and reducing circuit complexity.

\subsubsection{Downstream-task Layer}
After $L$ layers of QGCLs, the final node embeddings $v_{i}^{(L)}$ are used for downstream-tasks. For node-level tasks, such as device power control, each node embedding is passed through a task-specific MLP to produce an individual output~$y_{i}$. For graph-level tasks, an addition global pooling operation is applied over all node embeddings to generate a single graph representation. This operation can be summation, mean, or a learnable attention mechanism. 

{\color{black}
In summary, the SQM-GNN is explicitly designed for structurally mimicking the classical architecture. However, quantum-domain operations further enhance structural learning and provide computational advantages. 
Here, following existing work on quantum-assisted wireless communications~\cite{QuantumComputation}, we quantifying computational complexity in terms of the numbers of cost-function evaluations (CFEs), where one CFE corresponds to a single message passing forward.
In SQM-GNN, each QGCL operates on a fixed-size star-shaped subgraph having at most $k$ neighbors. Thanks to quantum superposition, the underlying PQC processes the entire subgraph in a single step~\cite{QuantumComputation}. Thus, on a graph having $N$ nodes, a full QGCL forward pass requires $\mathcal{O}(N)$ CFEs, independent of the specific node degrees.
By contrast, in a classical message passing GNN, each node aggregates messages from all neighbors, and each contribution is evaluated sequentially. The per-layer complexity is therefore $\mathcal{O}(\bar{d}N)$, where $\bar{d}$ is the average node degree, which can be substantially higher than $\mathcal{O}(N)$ in dense wireless graphs.
Moreover, subgraphs are generated on-the-fly from the original adjacency structure, so no additional large-scale subgraph storage is required.
}

\section{An Example Use Case of Scalable Quantum Message Passing GNN in Wireless Communication}\label{sec:Sim}

\begin{figure*}[t]
    \centering
    \begin{subfigure}[t]{0.55\linewidth}
        \centering
        \includegraphics[height=4cm]{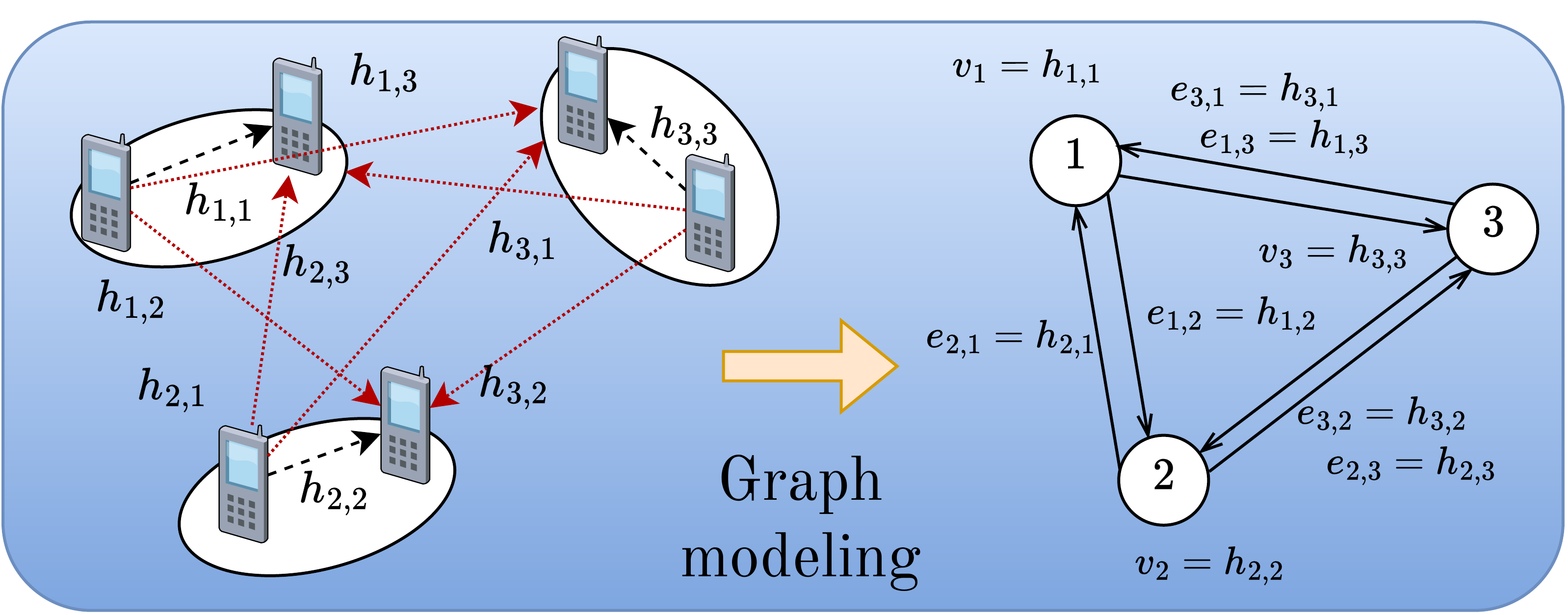}
        \caption{Graph-based representation of D2D system.}
        \label{fig:D2D_Model}
        \vspace{-0.1cm}
    \end{subfigure}
    \hspace{1cm}
    \begin{subfigure}[t]{0.35\linewidth}
        \centering
        \includegraphics[height=4cm]{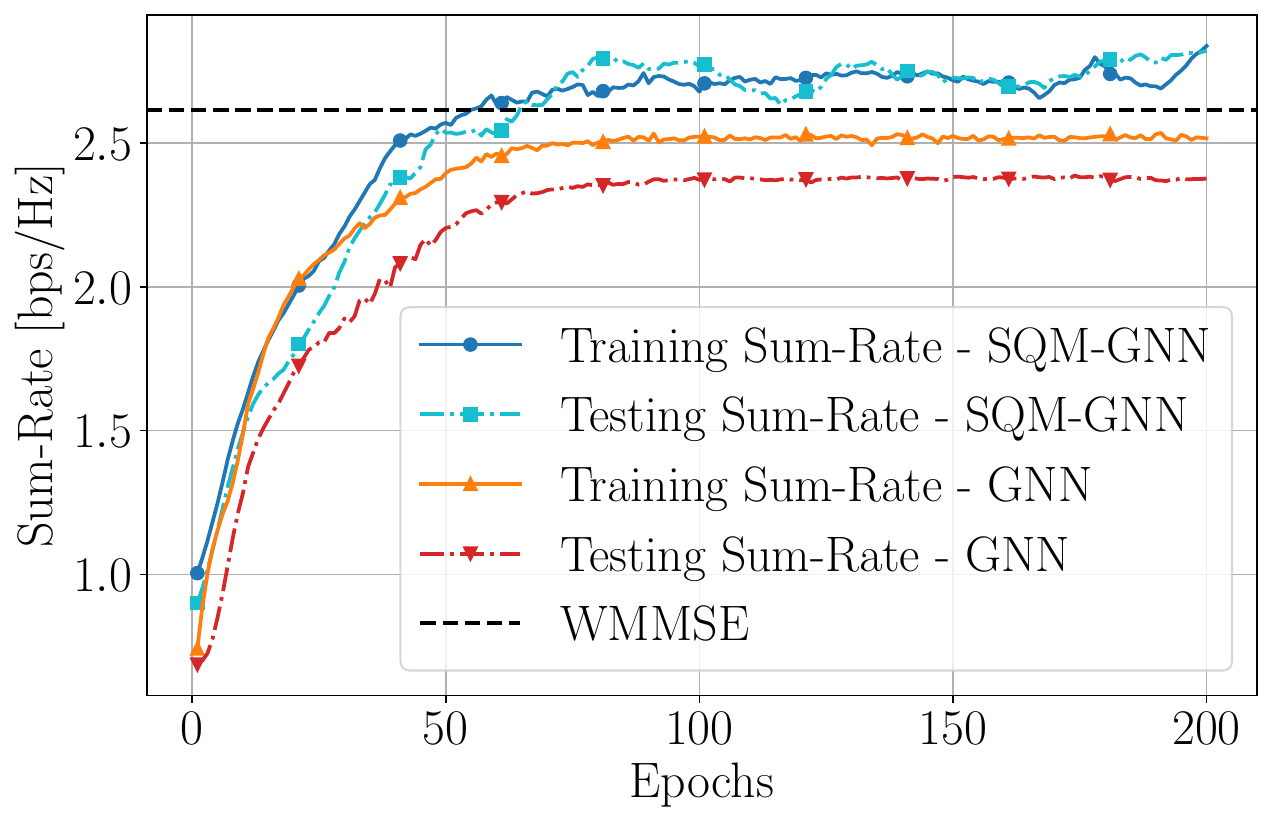}
        \caption{Sum-rate performance comparison.}
        \label{fig:D2D_Res}
        \vspace{-0.1cm}
    \end{subfigure}
    \centering
    \caption{SQM-GNN application of power control in a D2D network.}
    \label{fig:SQM-GNN_D2D}
    \vspace{-0.5cm}
\end{figure*}



In this section, we illustrate the application of the proposed SQM-GNN to the power control problem of an interference-limited D2D system.
A D2D communication system can be naturally represented as a graph, where each transmitter-receiver pair forms a node, and the edges represent mutual interference between pairs. The feature of each node corresponds to its direct desired channel gain, while edge features capture the interference channel gains between different D2D links. This results in an undirected, homogeneous graph structure, as shown in Fig.~\ref{fig:D2D_Model}.
In this graph-based representation, the SQM-GNN is used for learning the power control coefficient for each D2D pair. The message passing mechanism operates on the graph for iteratively updating the node embeddings, which are then decoded to predict the transmit power levels. 

\subsection{Loss Function \& Model Training}

The learning objective of SQM-GNN is to maximize the overall system sum-rate. This is equivalently formulated as minimizing the negative sum-rate across all $K$ D2D links, leading to the following loss function
\begin{align}
\mathcal{L} = -\sum_{k=1}^{K}  \log_2 \left(1 + \frac{p_k h_{kk}}{\sum_{j \neq k} p_j h_{jk} + \sigma^2} \right),
\end{align}
where $p_k$ denotes the predicted transmit power of user $k$, $h_{kk}$ and $h_{jk}$ represent the direct and interfering channel gains, respectively, and $\sigma^2$ is the noise power.
Let $N_{\text{shot}}$ denote the number of PQC measurement shots required per subgraph, and $N$ be the number of nodes in the graph. For a mini-batch size of $B$ and $T$ training epochs, the total number of CFEs is given by $\mathcal{C}_{\text{train}}~=~T~\cdot~B~\cdot~N~\cdot~N_{\text{shot}}$.
Since the same PQC is shared across all subgraphs, $N_{\text{shot}}$ remains constant, thus, the training complexity scales linearly with the number of nodes, i.e., with the order of $\mathcal{O}(N)$.

\subsection{Simulation Results}

Simulations are conducted over a square area of $500~\times~ 500~\mathrm{m^2}$ having $K=20$ single-antenna D2D pairs. The maximum transmission power is set to $\bar{p}=1~\mathrm{ [W]}$, and the noise power is $-104~\mathrm{dBm}$. The channel gains follow a standard path-loss model similar to~\cite{2021_JSAC_GNN_Shen}. To evaluate the performance, we generate $10,000$ realizations for both training and testing. 
\textcolor{black}{
For benchmarking, we employ the popular WMMSE algorithm and the classical GNN model of~\cite{2021_JSAC_GNN_Shen}. 
In all experiments, WMMSE is run for $100$ iterations, which is sufficient to ensure convergence to a Karush-Kuhn-Tucker (KKT) stationary point for the D2D power allocation problem considered.
}
\textcolor{black}{
All simulations are implemented in Python. The proposed SQM-GNN is implemented as a hybrid classical-quantum model using PennyLane~\cite{pennylane}, which seamlessly provides differentiable PQC layers and interfaces with PyTorch. Each experiment is repeated five times with different random seeds, and we report the mean as well as standard deviation. All simulations are conducted on a workstation using an AMD Ryzen 9 7950X CPU and $128$GB of RAM.
}

\textcolor{black}{
As shown in Fig.~\ref{fig:D2D_Res}, the GNN converges in fewer training epochs, but remains sub-optimal relative to WMMSE. 
By contrast, SQM-GNN exploits a higher-dimensional quantum feature space and therefore requires more epochs. As a benefit, it ultimately learns better power allocation policies than WMMSE. 
After $100$~epochs, SQM-GNN achieves a testing sum-rate of approximately $2.6~\mathrm{bps/Hz}$, while the classical GNN reaches only around $2.3~\mathrm{bps/Hz}$. 
Notably, SQM-GNN is capable of finding a slightly better  stationary point than that provided by WMMSE and also exhibits better generalization than the classical GNN, as indicated by the smaller gap between its training and testing curves.
}

\begin{table}[t]
    \centering
    \renewcommand{\arraystretch}{0.9}
    \setlength{\tabcolsep}{0.5em} 
    \caption{Scalability and Generalization Capability Comparison of SQM-GNN and GNN.}
    \vspace{-0.2cm}
    \resizebox{\columnwidth}{!}{
    \begin{tabular}{cccc|cccc}
        \toprule
         \midrule
         $K$& $\bar{p}$ &GNN \cite{2021_JSAC_GNN_Shen} & SQM-GNN 
         &$K$ &$\bar{p}$ &GNN \cite{2021_JSAC_GNN_Shen} & SQM-GNN  \\
         \midrule
         $10$ &$1$ & $102.21\%$ & $107.02\%$ 
         &$10$ &$2$ & $102.00\%$ & $106.88\%$  \\
         $20$ &$1$ & $95.16\%$ & $105.12\%$ 
         &$20$ &$2$ & $95.99\%$ & $103.28\%$  \\
         $40$ &$1$ & $93.99\%$ & $100.59\%$ 
         &$40$ &$2$ & $93.74\%$ & $99.63\%$  \\
         $80$ &$1$ & $93.38\%$ & $97.87\%$ 
         &$80$ &$2$ & $92.01\%$ & $98.26\%$  \\
         \midrule
         \bottomrule
    \end{tabular}
    }
    \label{tab:res_scale}
    \vspace{-0.5cm}
\end{table}

Table~\ref{tab:res_scale} compares the scalability and generalization performance of the classical GNN and SQM-GNN models under various network sizes and transmit power constraints, expressed as a percentage of the WMMSE benchmark.
Both models are trained under a fixed setting of $K = 20$ and $\bar{p} = 1 \mbox{ [W]}$, and they are evaluated under different conditions without retraining. As shown, SQM-GNN achieves consistently better generalization across a wide range of $K\in\{10,20,30,40\}$ and $\bar{p}\in\{1,2\}\mbox{ [W]}$. For instance, when tested with $K = 10$ and $\bar{p} = 1\mbox{ [W]}$, SQM-GNN exceeds the WMMSE baseline with a score of $107.02\%$, while the classical GNN reaches $102.21\%$. As network size increases to $K = 80$ with power budget of $\bar{p}=2 \mbox{ [W]}$, SQM-GNN still performs well at $98.26\%$, whereas the classical GNN drops to $92.01\%$.
These results highlight the advantage of the PQC-based message passing design in SQM-GNN, which improves the model's robustness and scalability across different network sizes and power budgets. This is especially important in wireless networks, where system conditions frequently vary and generalization is critical.


\begin{table}[t]
    \color{black}
    \centering
    \renewcommand{\arraystretch}{0.85}
    \setlength{\tabcolsep}{0.6em}
    \caption{\color{black} Scalability and Feasibility Analysis of QSGCN, SQM-GNN, and GNN.}
    \vspace{-0.2cm}
    \begin{tabular}{lrcccccc}
        \toprule
        \midrule
        \multicolumn{2}{c}{Model} & $K=4$ & $K=8$ & $K=10$ & $K=20$ \\
        \midrule

        \multirow{2}{*}{QSGCN \cite{2025_QSGCN_TNNLS}} &\#qubits 
            & $10$ & $36$ & $55$ & $210$ \\
        &Feasibility
            & $\checkmark$ & $\checkmark$ & $\times$ & $\times$\\
        \midrule
        \multirow{2}{*}{\makecell[l]{SQM-GNN\\$(k=6)$}} &\#qubits
            & \multicolumn{4}{c}{$\mathbf{13}$} \\
        &\#Params
            & \multicolumn{4}{c}{$\mathbf{5,925}$} \\
        \midrule
        GNN \cite{2021_JSAC_GNN_Shen} &\#Params
            & \multicolumn{4}{c}{$\mathbf{67,073}$} \\
        \midrule
        \bottomrule
    \end{tabular}
    \label{tab:res_param}
    \vspace{-0.5cm}
\end{table}

\textcolor{black}{
Table~\ref{tab:res_param} compares the scalability of the QSGCN, GNN baseline, and the proposed SQM-GNN for varying network sizes $K$. 
In QSGCN, the number of qubits required grows quadratically with $K$, leading to excessive complexity for networks having $K \geq 10$ D2D pairs, which is far below practical scales. 
By contrast, thanks to the subgraph decomposition strategy, SQM-GNN preserves the inherent scalability of classical GNNs. 
SQM-GNN operates on fixed-size subgraphs, requiring only $13$ qubits for a subgraph size of $k=6$. 
Moreover, SQM-GNN is highly efficient, using only $5,925$ trainable parameters, which represents a $90\%$ reduction compared to the GNN model.
}

\section{Open Questions and Future Directions}\label{sec:Discuss}
In this section, we highlight the key limitations of the SQM-GNN framework and outline several promising research directions for advancing quantum-assisted GNN in NG systems.

\begin{table*}[t]
    \centering
    \renewcommand{\arraystretch}{1.1}
    \setlength{\tabcolsep}{0.5em} 
    \caption{Trade-off between large quantum circuits and small quantum circuits}
    \begin{tabular}{p{0.15\linewidth}@{\hspace{1em}}p{0.38\linewidth}@{\hspace{1em}}p{0.38\linewidth}}
        \toprule
         \midrule
         &\makecell[c]{Large quantum circuits} &\makecell[c]{Small quantum circuits}  \\
         \midrule
         Spatial dependencies & \textbf{Better:} Preserves more structural information & \textbf{Limited:} Information loss due to dimensionality reduction \\
        Expressivity & \textbf{High:} Richer entanglement structures & \textbf{Low:} Limited representational capacity \\
        Hardware feasibility & \textbf{Challenging:} Exceeds limitations of current NISQ devices & \textbf{Practical:} Feasible on available NISQ devices \\
        Quantum noise & \textbf{Higher:} Due to greater circuit depth and gate count & \textbf{Lower:} Fewer gates and shallower depth \\
        Trainability & \textbf{Difficult:} More prone to barren plateaus & \textbf{Easier:} Less susceptible to barren plateaus \\
         \midrule
         \bottomrule
    \end{tabular}
    \label{tab:qnn_tradeoff}
    \vspace{-0.5cm}
\end{table*}

\subsection{Open Questions}
Despite the promising results demonstrated by SQM-GNNs in wireless communication tasks, several open challenges should be addressed before they can be widely adopted in practical systems.

\subsubsection{Trade-off between Quantum Circuit Complexity and Subgraph Informativeness}
One of the most critical challenges in designing a SQM-GNN model lies in the trade-off between the size of the PQC and the amount of information aggregated from each subgraph. As discussed in Section~\ref{sec:Method}, each star-shaped subgraph is processed by a PQC, whose required number of qubits is proportional to the number of nodes and edges within the subgraph. However, current NISQ devices impose strict limitations on both qubit count and circuit depth, making this trade-off particularly important.

From the perspective of graph data information, fully capturing spatial dependencies requires each node to aggregate messages from all its neighbours, implying a larger subgraph. However, this increases circuit complexity and qubit requirements, leading to implementation infeasibility on practical quantum devices. 
Specifically, in more complex PQCs, gradients might vanish, degrading the trainability of QML models. By contrast, using smaller subgraphs allows for more feasible PQC execution, albeit at the cost of reduced contextual information in message aggregation. 

Table~\ref{tab:qnn_tradeoff} illustrates this trade-off between large and small quantum circuits. As shown, large circuits offer improved expressivity and better preservation of spatial structure but require more hardware resources and are more vulnerable to quantum noise and optimization difficulties. Conversely, small circuits are more suitable for current hardware and easier to train, yet their limited representational capacity often results in significant information loss. This underscores the need for careful circuit and subgraph design in SQM-GNNs for balancing information fidelity against quantum feasibility. For example, in the fully connected D2D graph described in Section~\ref{sec:Sim}, each device pair interferes with every other, hence sampling only a subset of neighbours leads to information loss in interference modeling. Consequently, the power control performance is severely degraded. This highlights a central challenge: how to design PQCs that remain hardware-efficient, while still capturing sufficient spatial correlations to support effective learning and decision-making.

\subsubsection{Efficient Quantum Data Representation}
While each qubit inherently represents information within a higher-dimensional Hilbert space compared to a classical bit, practical limitations of current quantum hardware significantly constrain the dimensionality of classical data that can be effectively encoded into quantum states. As a consequence, classical high-dimensional data must undergo dimensionality reduction before quantum encoding. According to information bottleneck theory \cite{2015_InformBottleneck}, such dimensionality compression inevitably introduces information loss. Without a carefully designed compression strategy, this information loss can significantly degrade the performance and learning capabilities of QML.

To address these challenges, it is essential to develop model-based or learning-based dimensionality reduction techniques under explicit information constraints, thereby preserving critical features of the original data. Alternatively, employing an encoder-decoder framework provides an effective solution,  where high-dimensional data can be reconstructed from compressed quantum representations. This can effectively alleviate the detrimental impact of information loss on the learning performance of quantum models.

\subsubsection{Quantum Circuit Design for Message Passing}
Another crucial challenge lies in the PQC design for the message passing mechanism. In contrast to traditional GNNs, where messages gleaned from neighboring nodes are aggregated simultaneously, message passing in PQCs must be explicitly encoded via sequential two-qubit operations. Additionally, according to quantum entanglement, superposition, and the no-cloning theorem, the quantum state changes after each gate operation. Consequently, the specific order of gates and operations critically affects the outcome of each PQC. Different entanglement topologies, for instance, may lead to varying degrees of information mixing, potentially introducing inductive bias, depending on the graph structure.

Moreover, as the depth and width of the PQC are increased for capturing richer entanglement, quantum noise becomes a critical concern. Accumulation of gate errors and decoherence over time degrades the fidelity of message transmission. Thus, the PQC must be carefully designed. In particular, currently, there are no standardized PQC design methodologies capable of balancing expressivity against information fidelity. This complexity positions circuit design as a central challenge in the development of quantum graph neural networks.

\subsection{Future Directions}
\subsubsection{Multi-Task Quantum Learning for Integrated NG Control}
NG networks are characterized by massive connectivity, dynamic topologies, and heterogeneous service demands. This necessitates learning architectures that go beyond single-task optimization. A promising direction involves extending the SQM-GNN architecture to a multi-task learning framework. 
In this design, the core sharing quantum message passing is retained for capturing both spatial and topological dependencies. 
In addition to this shared structure, task-specific readout heads are incorporated for simultaneously learning multiple control objectives. This includes power control, beamforming, user association, UAV trajectory planning, and access point activation. This approach aligns with the vision of QPU-assisted NG gateways, wherein a single quantum model operates as a unified controller across layers, domains, and services.

\subsubsection{Privacy-Preserving Communication}
As NG networks integrate user-driven services and mission-critical applications, privacy becomes a central concern. 
Conventional wireless optimization often relies on centralized processing, where a controller collects global information, which is problematic when devices generate sensitive or proprietary data. 
Federated learning (FL) offers an attractive alternative by training models locally and sharing only model updates. Combining FL with SQM-GNN would enable decentralized, privacy-preserving quantum-assisted learning while retaining scalability. Moreover, the modular nature of FL fits well with distributed and hierarchical 6G architectures, making federated SQM-GNN a promising direction for privacy-aware, secure quantum-aided NG communications.

\subsubsection{Cross-Domain Adaptation and Meta-Learning}
The architecture of 6G networks will span terrestrial, aerial, and non-terrestrial domains, each with its own constraints and characteristics.
Despite these differences, they share a structural similarity in the form of graph-based connectivity. 
A critical direction for future research is enabling SQM-GNNs to generalize across these domains without requiring retraining or reconfiguration for every environment. This can be achieved through meta-learning strategies and domain adaptation techniques, allowing the quantum model to learn shared graph features that transfer across scenarios.
This cross-domain capability would support the broader goal of building agile, resource-efficient quantum learning systems capable of meeting the dynamic needs of NG infrastructures.

\section{Lesson Learned and Conclusions}
{\color{black}
\subsection{Lesson Learned}
Based on the proposed architecture and numerical evaluations, our proposition of integrating quantum computing and GNNs in NG wireless systems yields some important insights:
\begin{itemize}[leftmargin=0.5em, itemsep=0.1em, topsep=0.1em]
    \item While GNNs show promise for wireless communication problems thanks to their scalability and strong generalization capability, they may suffer from a computational bottleneck in classical processors. As a remedy, hybrid quantum-classical GNN architectures present a promising path forward, but, naively replacing classical MLP by a QNN yields limited benefits. 
    \item Quantum advances can be beneficially leveraged in GNNs, when the message passing operations are embedded directly into PQCs, enabling expressive quantum representations of relational information. However, NISQ-era hardware limitations severely restrict qubit counts and gate fidelities, hence preventing full-graph based processing on current devices. A practical solution is to decompose the graph into subgraphs relying on limited quantum resources. However, this requires careful design for ensuring flawless message passing and preserve effective inter-subgraph connectivity.
    \item Naturally, this subgraph decomposition philosophy imposes an inevitable trade-off between the hardware resources and the aggregated information exchanged amongst nodes. Smaller subgraphs accommodate hardware limitations but restrict how much neighbourhood information each node can aggregate. Balancing qubit availability against the expressiveness of message passing remains a pivotal design challenge that must be carefully investigated.
\end{itemize}

\subsection{Conclusions}
This article introduced SQM-GNN, a scalable quantum graph neural network based technique conceived for NG wireless communication, where message passing is implemented via PQCs. Quantum-enhanced message passing allows SQM-GNN to capture more expressive structural dependencies in wireless graphs, while subgraph decomposition preserves the scalability and generalization properties of classical GNNs under NISQ constraints.
We characterized the benefits of SQM-GNNs in a D2D power control context and observed both superior performance and strong generalization compared to both classical GNN baselines and existing quantum-native GNNs, indicating that quantum-aided message passing facilitates topology-aware learning for future wireless systems. 
As quantum hardware matures, SQM-GNN offers a promising basis for intelligent, real-time control in NG networks, with future directions including cross-domain adaptation, multi-task integration, and privacy-preserving quantum learning.
}

\bibliographystyle{IEEEtran}
\bibliography{reference}

\begin{thebibliography}{10}
\providecommand{\url}[1]{#1}
\csname url@samestyle\endcsname
\providecommand{\newblock}{\relax}
\providecommand{\bibinfo}[2]{#2}
\providecommand{\BIBentrySTDinterwordspacing}{\spaceskip=0pt\relax}
\providecommand{\BIBentryALTinterwordstretchfactor}{4}
\providecommand{\BIBentryALTinterwordspacing}{\spaceskip=\fontdimen2\font plus
\BIBentryALTinterwordstretchfactor\fontdimen3\font minus
  \fontdimen4\font\relax}
\providecommand{\BIBforeignlanguage}[2]{{%
\expandafter\ifx\csname l@#1\endcsname\relax
\typeout{** WARNING: IEEEtran.bst: No hyphenation pattern has been}%
\typeout{** loaded for the language `#1'. Using the pattern for}%
\typeout{** the default language instead.}%
\else
\language=\csname l@#1\endcsname
\fi
#2}}
\providecommand{\BIBdecl}{\relax}
\BIBdecl

\bibitem{2024_Tung_GNNSurvey}
N.~X. Tung \emph{et~al.}, ``{Graph Neural Networks for Next-Generation-IoT:
  Recent Advances and Open Challenges},'' \emph{IEEE Commun. Surveys Tuts.},
  2025.

\bibitem{2024_SensJ_Giang_HGNN}
L.~T. Giang \emph{et~al.}, ``{{HGNN}: A Hierarchical Graph Neural Network
  Architecture for Joint Resource Management in Dynamic Wireless Sensor
  Networks},'' \emph{IEEE Sensors J.}, vol.~24, no.~24, pp. 42\,352--42\,364,
  2024.

\bibitem{2024_TVT_Tung_GNN_cfmMIMO}
N.~X. Tung \emph{et~al.}, ``{Distributed Graph Neural Network Design for Sum
  Ergodic Spectral Efficiency Maximization in Cell-Free Massive MIMO},''
  \emph{IEEE Trans. Veh. Technol.}, vol.~74, no.~3, pp. 5181--5186, 2025.

\bibitem{2017_NIPS_GraphSAGE}
W.~L. Hamilton \emph{et~al.}, ``Inductive representation learning on large
  graphs,'' in \emph{Proc. 31st Int. Conf. Neural Inf. Process. Syst. (NIPS)},
  2017, p. 1025–1035.

\bibitem{2024_TWC_QML}
J.~A. Ansere \emph{et~al.}, ``{Quantum Deep Reinforcement Learning for Dynamic
  Resource Allocation in Mobile Edge Computing-Based IoT Systems},'' \emph{IEEE
  Trans. Wirel. Commun.}, vol.~23, no.~6, pp. 6221--6233, 2024.

\bibitem{2025_QSGCN_TNNLS}
J.~Zheng \emph{et~al.}, ``{A Quantum Spatial Graph Convolutional Neural Network
  Model on Quantum Circuits},'' \emph{IEEE Trans. Neural Netw. Learn. Syst.},
  vol.~36, no.~3, pp. 5706--5720, 2025.

\bibitem{II_B_2021_TPDS_Singh_A_QComp}
A.~K. Singh \emph{et~al.}, ``{A Quantum Approach Towards the Adaptive
  Prediction of Cloud Workloads},'' \emph{IEEE Trans. Parallel Distrib. Syst.},
  vol.~32, no.~12, pp. 2893--2905, 2021.

\bibitem{2025_Hieu_QCNN_cfmMIMO}
D.~H. Nguyen \emph{et~al.}, ``{Hybrid Quantum Convolutional Neural
  Network-Aided Pilot Assignment in Cell-Free Massive MIMO Systems},''
  \emph{IEEE Trans. Veh. Technol.}, 2025.

\bibitem{2018_NISQ_John}
J.~Preskill, ``Quantum {C}omputing in the {NISQ} era and beyond,''
  \emph{{Quantum}}, vol.~2, p.~79, Aug. 2018.

\bibitem{II_B_Plateau_Lucas_F}
L.~Friedrich and J.~Maziero, ``{Evolution strategies: application in hybrid
  quantum-classical neural networks},'' \emph{Quantum Inf. Process.}, vol.~22,
  no.~3, p. 132, 2023.

\bibitem{2021_PhysRev_Schuld_Encoding}
M.~Schuld \emph{et~al.}, ``Effect of data encoding on the expressive power of
  variational quantum-machine-learning models,'' \emph{Phys. Rev. A}, vol. 103,
  p. 032430, Mar 2021.

\bibitem{QuantumComputation}
P.~Botsinis \emph{et~al.}, ``{Quantum Search Algorithms for Wireless
  Communications},'' \emph{IEEE Commun. Surveys Tuts.}, vol.~21, no.~2, pp.
  1209--1242, 2019.

\bibitem{2021_JSAC_GNN_Shen}
Y.~Shen \emph{et~al.}, ``{Graph Neural Networks for Scalable Radio Resource
  Management: Architecture Design and Theoretical Analysis},'' \emph{IEEE Jour.
  of Sel. Areas in Comm.}, vol.~39, no.~1, pp. 101--115, 2021.

\bibitem{pennylane}
V.~Bergholm \emph{et~al.}, ``Pennylane: Automatic differentiation of hybrid
  quantum-classical computations,'' \emph{arXiv:1811.04968[quant-ph]}, 2022.

\bibitem{2015_InformBottleneck}
N.~Tishby and N.~Zaslavsky, ``{Deep Learning and the Information Bottleneck
  Principle},'' \emph{arXiv:1503.02406[cs.LG]}, 2015.

\end{thebibliography}

\end{document}